\documentclass{pasa}

\usepackage{graphicx}
\usepackage{amsmath}
\usepackage{aas_macros}
\usepackage{amssymb}
\usepackage{hyperref}

\newcommand{\be}{\begin{equation}}
\newcommand{\ee}{\end{equation}}
\newcommand{\beq}{\begin{eqnarray}}
\newcommand{\eeq}{\end{eqnarray}}
\newcommand{\n}{\mathrm{n}}
\newcommand{\p}{\mathrm{p}}
\newcommand{\xx}{\mathrm{x}}
\newcommand{\y}{\mathrm{y}}
\newcommand{\vv}{\mathrm{v}}
\newcommand{\R}{\mathcal{R}}

\title[Hydrodynamics of vortex avalanches]{Modelling pulsar glitches: the hydrodynamics of superfluid vortex avalanches in neutron stars.}

\author[V.Khomenko \& B.Haskell]{
	V.~Khomenko and B.~Haskell
\affil{Nicolaus Copernicus Astronomical Center, Polish Academy of Sciences, ul. Bartycka 18, 00-716 Warsaw, Poland}
}

\jid{PASA}
\doi{10.1017/pas.\the\year.xxx}
\jyear{\the\year}

\begin{document}

\begin{frontmatter}
\maketitle

\begin{abstract}

The dynamics of quantised vorticity in neutron star interiors is at the heart of most pulsar glitch models. However the large number of vortices (up to $\approx 10^{13}$)  involved in a glitch and the huge disparity in scales between the femtometre scale of vortex cores and the kilometre scale of the star makes quantum dynamical simulations of the problem computationally intractable. In this paper we take a first step towards developing a mean field prescription to include the dynamics of vortices in large scale hydrodynamical simulations of superfluid neutron stars. We consider a one dimensional setup and show that vortex accumulation and differential rotation in the neutron superfluid lead to propagating waves, or `avalanches', as solutions for the equations of motion for the superfluid velocities. We introduce an additional variable, the fraction of free vortices, and test different prescriptions for its advection with the superfluid flow. We find that the new terms lead to solutions with a linear component in the rise of a glitch, and that, in specific setups, they can give rise to glitch precursors and even to decreases in frequency, or `anti-glitches'.

\end{abstract}


\end{frontmatter}

\section{Introduction}

Pulsar glitches are sudden increases in the rotational frequency of pulsars (predominantly observed in radio, but also seen in X-rays and gamma-rays), that are instantaneous to the accuracy of the data (with the best upper limits constraining the rise time to be less than approximately one minute \citep{Rise}) and are thought to be due to the presence of a large scale superfluid component in the neutron star.
Mature neutron stars are, in fact, cold enough for neutrons to be superfluid and protons superconducting \citep{Migdal59, Baym69} (see \citet{HasSed2017} for a recent review). Superfluidity has a strong impact on the dynamics of the stellar interior, as a superfluid rotates by forming an array of quantised vortices which carry the circulation and mediate angular momentum exchange between the superfluid neutrons and the normal component of the star, which is tracked by the electromagnetic emission. 

If vortices are strongly attracted, or `pinned' to ions in the crust or flux tubes in the core of the star they cannot move out, and the superfluid cannot spin-down with the normal component, thus storing angular momentum. Sudden recoupling of the components leads to rapid angular momentum exchange and a glitch \citep{AndItoh75}.
Realistic models of pinning forces \citep{Seveso16, Wlaz16} and effective masses \citep{ChamelMass, Chamel17, Watanabe17} in the neutron star crusts can be used to calculate the amount of angular momentum transferred during a glitch \citep{Pierre11, Has12, Seveso12, Crust12, Chamel13} and compared to observations to constrain the mass of a glitching pulsar and its equation of state \citep{Newton15, Ho15, MassPierre}. Nevertheless the trigger mechanism for glitches is still unknown (see \citet{HasRev} for a recent review). The main mechanisms that have been proposed are crust-quakes \citep{Rud69}, hydrodynamical instabilities \citep{MM05, Kostas09} and vortex avalanches \citep{Cheng88, Lila13}.

In this paper we will focus on this last mechanism, vortex avalanches. In this model local, fast, interactions between neighbouring vortices release stresses built up gradually over time as the star spins down. This is the hallmark of Self Organised Criticality (SOC), and in fact quantum mechanical simulations of vortices in a spinning down trap confirm that the spin-down of the superfluid occurs via discrete vortex avalanches, and that the distribution of glitch sizes is a power-law, and the distribution of waiting times an exponential \citep{Lila11}. This is consistent with the size and waiting time distributions of most pulsars, except for Vela and J0537-6910 \citep{Melatos08}, for which an excess of large avalanches appears, with a preferred size and waiting time. Furthermore in J0537-6910 there is a correlation between waiting times and the size of the previous glitch \citep{Middle06, Ant17}, which some have suggested may be due to crust-quakes \citep{Middle06}, but may also be the consequence of rapid driving preventing the system from self-organising, and leading to the whole pinned vorticity being expelled once the maximum of the pinning force is reached, as is the case in the `snowplow' model for giant glitches \citep{Pierre11}. An excess of large avalanches may also indicate a departure of the system from SOC behaviour, and the onset of a self organised bistable state \citep{bistabile}.

Quantum mechanical simulations, however, suffer from numerical limitations that do not allow to simulate the full neutron star system, or to model the difference in scales in a neutron star, where vortices with a coherence length $\xi_c\approx 10$ fm are separated by a distance $d_\vv\approx 10^{-3}$ cm. Recent work has, however, shown that even in a realistic neutron star, vortices can always move, on average, far enough to knock on neighbouring vortices without re-pinning, provided the relative velocity between the normal fluid and the superfluid $W$ is close to critical velocity for unpinning, $W_{cr}$, and in particular $W\gtrsim 0.95 W_{cr}$ for standard superfluid drag parameters \citep{Haskellhop}. 

In this paper we take a first step towards including the microphysics of vortex avalanches in larger scale hydrodynamical simulations. The situation is complex, as for a hydrodynamical treatment one has to average over several vortices to define a coarse-grained momentum for the superfluid condensate, thus losing information on the dynamics of individual vortex unpinning and knock-on events. Large scale hydrodynamical coupling, however, can have a strong impact on the dynamics of the system, and it is well known since the pioneering work on vortex creep of \citet{Alpar84a} that thermal unpinning of vortices can lead to non linear terms in the hydrodynamical equations of motion for the superfluid velocities, and affect the post-glitch relaxation \citep{Akbal17}.

Recently \citet{Haskell16} showed that random unpinning events, drawn from micro-physically motivated power law distributions, can be included in simulations. The resulting glitch distribution, however, differs from the original distribution of vortex unpinning events, and has a cut-off at small sizes, which can explain the observed deviation from a power-law of the distribution of glitches in the Crab pulsar for small sizes \cite{EspMin}. Furthermore such an approach can also explain the different kinds of relaxation observed after glitches, also in the same star \citep{HasAnt}.
Here we will follow this approach, but take a step forward in modelling the non-linear propagation of a vortex unpinning front during an avalanche.

\section{Superfluid hydrodynamics}

We take as our starting point a hydrodynamical description of the superfluid interior of the star, in which we do not deal with the dynamics of individual vortices directly, but rather deal with averaged large scale degrees of freedom that describe superfluid neutron velocities and densities, and those of a charge neutral fluid of protons and electrons. For the purpose of investigating the timescales associated with glitches we will consider protons and electrons to be locked by electromagnetic interactions and flow as a single fluid \citep{Mendell91a}.
Following \citet{AndComer} we can write conservation laws for each species:
\be
\partial_t\rho_\xx+\nabla_i(\rho_\xx v_\xx^i)=0,
\ee
where the constituent index $\xx$ labels either protons ($\p$) or neutrons ($\n$), ${v}_{\xx}^{i}$ is the velocity and ${\rho}_{\xx}$ is a density of respective constituent $\xx$. Note also that summation over repeated indices is implied (with the exclusion of constituent indices). The Euler equations are:
\beq
&&(\partial_t+v_\xx^j\nabla_j)(v_i^\xx+\varepsilon_\xx w_i^{\y\xx})+
\nabla_i(\tilde{\mu}_\xx+\Phi) + 
\nonumber
\\ %
&&  +\varepsilon_\xx w^j_{\y\xx}\nabla_i v^\xx_j=(f_i^\xx+f_i^{\xx\p})/\rho_\xx,
\eeq
where $w_i^{\y\xx}=v_i^\y-v_i^\xx$ and $\tilde{\mu}_\xx=\mu_\xx/m_\xx$ is the chemical potential per unit mass (in the following we take $m_\p=m_\n$). The gravitational potential is $\Phi$ and $\varepsilon_\xx$ is the entrainment coefficient which can account for the reduced mobility of neutrons, especially in the crust \citep{Prix04, Chamel17}.
The terms on the right hand side are the contribution to the vortex-mediated mutual friction due to pinned vortices $f_i^{p\xx}$ and to free vortices, $f_i^{\xx}$, which, for straight vortices and laminar flows, takes the form:
\be
f_i^{\xx}=\kappa n_\vv \rho_\n \mathcal{B}^{'}\epsilon_{ijk}\hat{\Omega}^i_\n w^k_{\xx\y}+\kappa n_\vv\rho_\n\mathcal{B}\epsilon_{ijk}\hat{\Omega}^j_\n\epsilon^{klm}\hat{\Omega}^\n_l w_m^{\xx\y},
\label{MF}\ee
where $\Omega_\n^j$ is the angular velocity of the neutrons (a hat represents a unit vector), $\kappa=h/2m_\n$ is the quantum of circulation and $n_\vv$ is the vortex density per unit area, $\epsilon_{ijk}$ is the Levi Civita symbol. Finally the Feynman relations link vortex density at a cylindrical radius $\varpi$ to the rotation rate of a superfluid element:
\be
\kappa n_\vv (\varpi)=2\tilde{\Omega}_\n+\varpi \frac{\partial \tilde{\Omega}_\n}{\partial \varpi}, 
\ee
with 
\be
\tilde{\Omega}_\n=\left[\Omega_\n + \varepsilon_\n (\Omega_\p - \Omega_\n)\right].
\ee
The parameters $\mathcal{B}$ and $\mathcal{B}^{'}$ in the expression for the mutual friction in (\ref{MF}) can be expressed in terms of a dimensionless drag parameter $\mathcal{R}$, related to the standard drag parameter $\gamma_d$ as
\be
\mathcal{R}=\frac{\gamma_d}{\kappa \rho_\n},
\ee
and are defined as
\be
\mathcal{B}=\frac{\R}{1+\R^2}\;\;\;\;\mbox{and}\;\;\;\;\mathcal{B}^{'}=\frac{\R^2}{1+\R^2}.
\ee
The parameter $\R$ encodes the microphysics of the dissipation processes that take place in the stellar interior, and its value is highly uncertain. Nevertheless the exact value of $\R$ is not crucial for the following discussion, and we will in general assume that $\R\ll 1$, and indicate the values we use explicitly in the examples provided.

Following \citet{TrevGlitch} we can simplify the problem by considering the evolution of the angular velocity of two axially symmetric rotating components. The evolution equations for the angular velocities thus take the form: 
\beq
\dot{\Omega}_\n (\varpi)\!\!\!\!\!&=&\!\! \!\!\!\frac{Q(\varpi)}{\rho_\n}\frac{1}{1-\varepsilon_\n-\varepsilon_\p}+\frac{\varepsilon_\n}{(1-\varepsilon_\n)}\dot{\Omega}_{\mbox{ext}}+F_p
\label{eom1}
\\
\dot{\Omega}_\p (\varpi) \!\!\!\!\!&=&\!\!\!\!\!- \frac{Q(\varpi)}{\rho_\p}\frac{1}{1-\varepsilon_\n-\varepsilon_\p}-\dot{\Omega}_{\mbox{ext}} -\frac{\rho_\n}{\rho_\p} F_p \, ,
\label{eom2}
\eeq
where 
\be
Q(\varpi)=\rho_\n\gamma\kappa n_\vv \mathcal{B} (\Omega_\p-\Omega_\n),
\ee
and $\gamma=\n_f/n_\vv$ is the fraction of vortices which are not pinned (with $n_f$ the surface density of free vortices), while $\dot{\Omega}_{\mbox{ext}}$ is the contribution from the external spin-down torque. $F_p$ is the contribution from the pinning force. While recent progress has been made on determining the maximum value of the pinning force from microphysical calculations \citep{Seveso16, Wlaz16}, its exact form is not known. This does not, however, hinder our discussion, for which the exact form for $F_\p$ is not necessary. It will be sufficient to assume that $F_\p$ balances the contribution to the mutual friction from the $(1-\gamma) n_\vv$ pinned vortices, below a critical threshold for the lag $\Delta\Omega_C$. Above  $\Delta\Omega_C$ all vortices are free and $F_p=0$ (see e.g. \citet{Seveso16} for realistic estimates of the maximum lag the pinning force can sustain). In the following we will thus not explicitly consider the pinning force, but simply assume a value of the critical lag $\Delta\Omega_C$.

We can further simplify the problem by assuming, as in \citet{Has12} and \citet{Haskell16}, that the proton component, consisting of the elastic crust and tightly coupled protons and electrons in the core, is rigidly rotating. This is likely to be a good approximation on timescales longer than the elastic and Alfven timescales in the crust, and simplifies our problem considerably (although see \cite{VE14} for a discussion of the short-timescale dynamics that is neglected with this approximation). In this approximation the equation of motion for the protons can be obtained from (\ref{eom2}) and is the following:
\be
\dot{\Omega}_\p =-\dot{\Omega}_\infty -\int \frac{\varpi^2}{I_{\p}} \left[\frac{Q(\varpi)}{1-\varepsilon_\n-\varepsilon_\p}+\rho_\n F_p \right] dV,
\ee
where ${I}_{\p}$ is the moment of inertia of the charged component.

Note that the above equations assume that vortices are straight and the rotation profile of the neutron fluid is axisymmetric. This may not be the case in the presence of strong density dependent entrainment, as is expected in the crust \citep{ChamelMass}. \citet{AP17} have analysed this problem and proposed a formalism that allows one to treat the problem as axially symmetric. Given that the equations are formally equivalent, and we do not consider a density dependent entrainment profile, we will ignore this complication in the following and continue working with the above set of equations which allow for a more transparent interpretation of the results.

It is also expected that turbulence may develop in neutron star interiors, leading to a turbulent polarized tangle of vortices and additional non-linear terms in the mutual friction force \citep{ASC07}.
The importance of turbulence can be assessed by examining the evolution equation for the vorticity:
\beq
\frac{\partial \xi_{\n}^{k}}{\partial t}&=&(1-\mathcal{B}^{'})\epsilon^{kim}\nabla_m (\epsilon_{ijl}v_{\n}^{j}\xi _{\n}^{l})\nonumber\\
&&+\mathcal{B}{{\epsilon^{kim}} }\nabla_m (\xi^{\n}_{i}\hat{\xi}^{\n}_{j} v_{\n}^{j}-\hat{\xi}^{\n}_{j}\xi_{\n}^{j}v_{\n}^{i})\;,
\eeq
with $\xi_{i}^{\n} = {\epsilon}^{ijk}{\nabla}_{j }v_{k}^{\n}$, and where $\hat{\xi}_i$ is a unit vector along $\xi_i$. 

The first term on the right hand side represents transfer of energy to small length scales, while the second leads to damping that stabilises the flow. The relative importance of two effects is determined by a parameter:
\be
q=\frac{\mathcal{B}}{1-\mathcal{B}^{'}}.
\ee
It was shown by \citet{Finne03} that turbulence sets in for $q\leq 1.3$. At high densities in the stellar interior $ \mathcal{B}' \approx {\mathcal{B}}^{2} \ll 1$ so that $ q \ll 1 $ and a superfluid neutron core is expected to be extremely susceptible to becoming turbulent, while the importance of turbulence in the crustal region is more uncertain, given the large uncertainties on $\mathcal{B}$ and $\mathcal{B}^{'}$. To take into account the influence of turbulence and vortex curvature on the mutual friction force one may rewrite:
\beq
{{f}_{i}}^{mf}={\rho }_{\n}{L}_{R}({\mathcal{B}^{'}{\epsilon }_{ijk}{k}^{j}{{w}_{\n\p}}^{k}}+\mathcal{B}{\epsilon }_{ijk}{\epsilon }^{klm}{\hat{\kappa}}^{j}{\kappa}_{l}{w}_{m}^{\n\p}\nonumber\\
-\tilde{\nu} [\mathcal{B}^{'}{\hat{\kappa}}^{j}{\nabla}_{j}{\kappa}_{i}+\mathcal{B}{\epsilon}_{ijk}{\kappa}^{j}{\hat{\kappa}}^{l}]{\nabla}_{l}{\hat{\kappa}}^{k})+\frac{2{L}_{T}}{3}{\rho }_{\n}k \mathcal{B} {{w}^{\p\n}_{i}}\;,
\eeq
where the last term represents polarised turbulence while the third and forth describe the influence of an isotropic turbulent tangle. Here
\be
\tilde{\nu }=\frac{1}{1-{\varepsilon}_{\n}-{\varepsilon}_{\p}}\nu\;, 
\ee
with ${L}_{R}$ is the vortex length due to the rotation, such that $\left|\nabla\times {\vec{v}}_{\n} \right|=2\Omega ={L}_{R}\kappa$, and ${L}_{T} = L - {L}_{R}$, with $L$ the total length of vortex. The vector $\kappa^i$ defines the orientation of vortex and $\nu$ is a parameter that determines the tension of the vortex (see \citet{ASC07} for a detailed description of the problem).

For polarised turbulence the straight vortex term thus leads to the shortest timescales. In our work we study the short-timescale dynamics of avalanches and ignore longer post-glitch timescale, for which additional physics regarding coupling timescales at different densities would, anyway, have to be included (see, e.g. \citet{Has12} and \citet{Newton15}). Hence turbulence is unlikely to make a qualitative difference to our results, and for computational simplicity we ignore it in the following.

\section{Vortex pinning and unpinning}

To solve the system of equations in the previous section we still require inputs from microphysics. Apart from the fraction of neutrons, the mutual friction parameters $\mathcal{B}$ are required. In the outer-core of the neutron star the main contribution to mutual friction is expected to come from electrons scattering on vortex cores, which leads to $\mathcal{B}\approx 10^{-4}$ \citep{AlparMF, AndSid06}. In the crust the situation is much more uncertain, as phonon scattering will lead to weak damping, with mutual friction parameters as low as $\mathcal{B}\approx 10^{-10}$, but if vortices move rapidly past pinning sites, Kelvin waves may be excited leading to $\mathcal{B}\approx 10^{-2}$ \citep{JonesPhonon, JonesKelvon, EB92}. Given this level of uncertainty, and the small scales we consider, we will take $\mathcal{B}$ as a constant, free parameter, and study how our results depend on it.

In our description, however, we have also introduced an extra parameter that rescales the mutual friction coefficients, i.e. the fraction of unpinned vortices $\gamma$. This will depend specifically on the dynamics of vortices on scales smaller than the hydrodynamical scale we are discussing.


Before moving on let us thus address the validity of our hydrodynamical description. Hydrodynamics is the natural tool to model macroscopic, observable, phenomena in neutron stars, and the key assumption is that we can track the evolution of fluid elements as they evolve. A fluid element must be small enough to be considered as a `point' in the macroscopical hydrodynamical description, but also large enough to contain enough particles to allow for meaningful averaged hydrodynamical quantities and fluxes to be defined. 

This is particularly relevant for a superfluid, which is irrotational and rotates by forming an array of quantised vortices which carry the circulation. In practice this means that a coarse-grained description must average over several vortices in order to define a superfluid velocity for a fluid element, leading to a large scale neutron fluid which is not irrotational.
The minimum scale on which it is meaningful to discuss hydrodynamics is thus given by the typical inter-vortex spacing, which is of the order of:
\be
d_\vv=1\times 10^{-3} \left(\frac{P}{10 \mbox{ms}}\right)^{1/2} \mbox{cm},
\ee
where $P$ is the spin period of the star. Note, however that dynamics on a smaller inter-vortex scale (on which neutrons behave as an irrotational fluid, defined by the neutron-neutron scattering length-scale of approximately a micron), is crucial for determining mutual friction and pinning parameters.

Knock-on effects between vortices are likely to be fundamental for the dynamics we observe in pulsar glitches. Consider a simple model in which such effects are neglected, and one simply has random, uncorrelated, unpinning of individual vortices. In this case the probability of unpinning $n$ vortices during an observation time $\Delta t$, is simply a Poissonian \citep{Lila13}
\be
p(n)=\exp(-\theta\Delta t)\frac{(\theta \Delta t)^n}{n!}\label{poisson}\;,
\ee
where $\theta$ is the unpinning rate for a single vortex. The result in (\ref{poisson}) tends to a Gaussian for large $\Delta t$, suggesting that the average number of free vortices, and thus the average glitch size, should also follow a Gaussian distribution. However this conflicts with the data in all pulsars except for PSR J0537-6910 and the Vela pulsar \citep{Melatos08}. The glitch size distribution in other pulsars is consistent with a power-law, with the waiting times being exponentially distributed, which is in agreement with the results obtained with small scale quantum mechanical Gross Pitaevskii simulations of superfluid vortices in a spinning down trap \citep{Lila11}.
It is clear that we need to understand how to scale up the dynamics observed in such simulations of $\approx 10^2-10^3$ vortices, to the larger scale corse-grained hydrodynamical description where individual vortices are not resolved. 

On the one side the kind of vortex avalanches that are observed in simulations over scales of hundreds or thousands of vortices can lead to vortex depletion and accumulation on small scales and create sharp gradients in hydrodynamical simulations. On the other hand large scale dynamics can induce differential rotation, and significantly increase the local density of vortices, leading to knock on effects and unpinning \citep{Lila12unpin, Haskellhop}.
For example if vortex avalanches are free to propagate outward from high density regions at the base of the crust, towards lower density regions in which the pinning force peaks, individual vortice will encounter stronger pinning forces than in the region where they originally unpinned. This can lead to vortices accumulating close to the maximum of the pinning force. A similar situation may occur even if vortices creep out, but the rate is not fast enough to keep up with the external driver (the electromagnetic spin-down), leading to vortex accumulation \citep{Lila13}. This would create a vortex 'sheet' such as that suggested by \citet{Pierre11}, in which the vortex density is significantly higher than the steady state density $n_\vv=2\Omega_\n/\kappa$. Such a vortex sheet will gradually shift out, until the maximum lag that the pinning force can sustain is exceeded, after which all vortices are free and will rapidly move out transferring angular momentum catastrophically and giving rise to a `giant' glitch, such as those observed in the Vela pulsar.

\section{Vortex sheet}

Let us now discuss this scenario in more detail and study, from a hydrodynamical point of view how an avalanche can lead to angular momentum exchange between the superfluid and the crust.

First of all let us define the lag between the two components:
\be
\Delta\Omega=\Omega_\mathrm{p}-\Omega_\mathrm{n}\;.
\ee
By combining equations (\ref{eom1}) and (\ref{eom2}) we can obtain the following evolution equation
\be
\frac{\partial \Delta\Omega}{\partial t}=-\kappa n_\vv \frac{\gamma\mathcal{B}}{x_\p(1-\varepsilon_\n-\varepsilon_\p)}\Delta\Omega\;,\label{eom33}
\ee
where $x_\p=\rho_\p/\rho$ with $\rho=\rho_\p+\rho_\n$, and we have assumed that vortices are free ($F_p=0$). We ignore the external torque, as we will be studying dynamics on much shorter timescales than those on which the external spin-down is relevant.

In the standard case one neglects differential rotation and assumes that $\kappa n_v\approx 2\Omega_{\n}$. Neglecting the small change in overall frequency, and taking $\Omega_\n$ and $\mathcal{B}$ to be a constant, the approximate solution for the lag between the two components $\Delta\Omega$, is a damped exponential of the form \citep{AndSid06}:
\be
\Delta\Omega\approx \Delta_0 \exp{-(t/\tau)}\label{sol1}\;,
\ee
with $\Delta_0$ a constant and 
\be
\tau=\frac{x_{\p}(1-\varepsilon_{\n}-\varepsilon_{\p})}{2\Omega_{\n} \gamma \mathcal{B}}\;,\label{tscale}
\ee
with $\gamma$ taken to be constant. This is the mutual friction timescale that is usually compared to exponentially relaxing components of the spin frequency observed after glitches.

We intend to investigate a different situation here. We continue to consider the $\gamma=$constant case, and follow \citet{Cheng88}, thus considering a vortex accumulation region in which a large number of vortices (compared to the steady state number present in the region) has re-pinned, i.e. such that
\be
\kappa n_v\approx \varpi \frac{\partial}{\partial \varpi}[\Omega_\mathrm{n}+\varepsilon_\mathrm{n}(\Omega_\mathrm{p}-\Omega_\mathrm{n})]\;,
\label{eqkappa} \ee
where the steady state contribution $\kappa n_\vv\approx 2 \Omega_\mathrm{n}$ is neglected and $\varpi=r\sin\theta$ is the cylindrical radius.

In this situation it is likely that strong pinning will force vortices to re-pin immediately and they will not be able to adjust to the lattice-equilibrium position that would be needed for this large increase in density. By the time the approximation $\varpi \partial{\Omega_\mathrm{n}}\partial{\varpi}>2\Omega_\mathrm{n}$ is satisfied one has a change in vortex density of order unity, corresponding to a similar change in vortex spacing $l_v\approx \sqrt{n_v}$. This leads to a large change in Magnus force as vortices get closer to each other, and the vortex sheet is thus very likely to unpin \citep{Lila13, Haskell16}.

In the presence of significant differential rotation the equation of motion for the lag $\Delta\Omega$ takes the following form:
\be
\frac{\partial\Delta\Omega (\varpi, t)}{\partial t}=\varpi \frac{\gamma\mathcal{B}(1-\varepsilon_{\n})}{x_\mathrm{p}(1-\varepsilon_{\n}-\varepsilon_{\p})}\Delta\Omega(\varpi,t) \frac{\partial{\Delta\Omega (\varpi, t)}}{\partial\varpi}\;,
\ee
where we have assumed that, at least locally, the proton fluid (possibly the crust in this case) is rigidly rotating so that $\partial_\varpi \Omega_\p=0$ and thus $\partial_\varpi \Delta\Omega=-\partial\varpi \Omega_\n$. If we make the further approximation that in the crust the amount by which the vortices move is small compared to $\varpi_s\approx 10^6$ cm, and thus consider $\varpi=\varpi_s$ constant, the equation above becomes
\be
\frac{\partial\Delta\Omega (\varpi, t)}{\partial t}=-\beta\Delta\Omega(\varpi,t) \frac{\partial{\Delta\Omega (\varpi, t)}}{\partial\varpi} \;,
\label{24}
\ee
with 
\be
\beta=\varpi_s \frac{\gamma\mathcal{B}}{x_\mathrm{p}(1-\varepsilon_{\n}-\varepsilon_{\p})}(\varepsilon_{\n}-1).
\ee
Finally with the substitution $\Delta^*=\beta \Delta\Omega $ we obtain as a general form of equation
\be
\frac{\partial\Delta^*(\varpi, t)}{\partial t}=-\Delta^*(\varpi,t) \frac{\partial{\Delta^*(\varpi, t)}}{\partial\varpi} .
\ee
This is a Burgers equation, and allows travelling waves as solutions. 
In particular let us consider the following initial conditions, corresponding to a large number of vortices accumulated close to the maximum of the pinning force, such that the lag is negligible prior to the vortex accumulation region, and approximately the maximum value $\Delta\Omega_M$ after an infinitesimally thin accumulation region located at $\varpi_0$:
\be
\Delta\Omega (t=0)=\Delta\Omega_M \Theta ( \varpi-\varpi_0)\;,
\ee
corresponding to (given that we take $\beta$ to be a constant)
\be
\Delta^{*} (t=0)=\Delta^{*}_M \Theta ( \varpi-\varpi_0)\;,
\ee
where $\Theta(x)$ is the Heaviside step function.  A solution to this equation which conserves the total number of vortices (in the approximation $\varpi_s$=constant) is a fan wave:
\beq
\Delta^{*}(t,\varpi)&=&\frac{\varpi-\varpi_0}{t}\;\;\;\mbox{for}\;\;\;\varpi<\varpi_F\\
\Delta^{*}(t,\varpi)&=&\Delta^{*}_M\;\;\;\;\mbox{for}\;\;\;\varpi\geq\varpi_F \;,\label{sol2}
\eeq
with $\varpi_F=v_F t$, with $v_F=\Delta^{*}_M$. For our physical variable $\Delta\Omega$ this corresponds to:
\beq
\Delta\Omega(t,\varpi)&=&\frac{\varpi-\varpi_0}{\beta t}\;\;\;\mbox{for}\;\;\;\varpi<\varpi_F\\
\Delta\Omega(t,\varpi)&=&\Delta\Omega_M\;\;\;\;\mbox{for}\;\;\;\varpi\geq\varpi_F \;,\label{sol2}
\eeq
where the position of the front $\varpi_F$ moves at a speed $v_F=\beta\Delta\Omega_M=\Delta\Omega_M\varpi_s \frac{\gamma\mathcal{B}}{x_\mathrm{p}(1-\varepsilon_{\n}-\varepsilon_{\p})}(\varepsilon_{\n}-1)$ (keeping in mind that in a pulsar the lag $\Delta\Omega$ and $\beta$ in the crust are negative). After the wave has travelled a distance $d\approx 10^2$ cm our approximation (
$\varpi \partial_\varpi
	\left[\Omega_\mathrm{n}+\varepsilon_\mathrm{n}(\Omega_\mathrm{p}-\Omega_\mathrm{n})\right]
\gg
2\Omega_\mathrm{n}$) breaks down and one has once again $\kappa n_v\approx 2\Omega$. Furthermore, as the l
ag is below the critical value, one can expect repinning. However the above analysis shows vortex accumulation can lead to rapid outward motion of vortices, on length scales much larger that the inter vortex motion, that may drive further avalanches as it travels through the medium.

The number of vortices in the front is (assuming $\varpi$ constant)
\beq
N_\vv&=&\int \frac{n_\vv}{\kappa} dS \approx 2\pi \int \frac{\varpi_s^2}{\kappa}\frac{\partial \Delta\Omega}{\partial\varpi} dr \nonumber\\
&=& \frac{\varpi_s^2}{\kappa} \Delta\Omega \bigg|^{\varpi_F}_{\varpi_0}\approx \frac{\Delta\Omega_M}{\kappa}\;,
\eeq
which is conserved by the solution.

Let us now consider the region upstream from the vortex accumulation region. Here there must be a depletion region in which the lag has to change from it's equilibrium value $\Delta\Omega_e$ to 0 over a short lengthscale. This can be determined by imposing that vortex density vanish, i.e. 
\newline
\be
\kappa n_\vv = 2\Omega_\n+\varpi\frac{\partial\Omega_\n}{\partial \varpi} = 0\;,
\ee
which gives, close to a point $\varpi_d$
\be
\Omega_\n=\Omega_0\left(\frac{\varpi_d}{\varpi}\right)^2.
\ee
Expanding around $\varpi_d$ in $\delta r=(\varpi-\varpi_0)$, we see that the lag decreases as
\be
\Delta\Omega=\Delta\Omega_0-2\Omega_0\frac{\delta r}{\varpi_0}.
\label{limit}
\ee
For typical parameters this gives a decrease in lag over a lengthscale $\delta r\approx 10-100$ cm, which is the same lengthscale over which our approximation is valid.

Nevertheless, to continue our initial analysis we will make the simplifying assumption that in this region the drop in lag can be approximated as a steep drop of the form
\be
\Delta\Omega=\Delta\Omega_M\Theta(\varpi_{d}-\varpi)\;,
\ee
but one should keep in mind that this is at the limit of validity for using equation (\ref{24}), given the estimate in (\ref{limit})
In this case, as vortex unpinning causes a perturbation in Magnus force due to the lack of vortices in their equilibrium positions, there will be a backward propagating unpinning avalanche. The solution corresponds to a forward moving shock that describes vortices moving out to fill the void, of the form
\be
\Delta\Omega=\Delta\Omega_M\Theta(\varpi_{d}-\varpi)\label{sol3}\;,
\ee
where the position of the shock is determined by $\varpi_{d}=v_S t$, with $v_S=\beta\Delta\Omega_M/2$.

\section{Unpinning vortex waves}

In the previous section we showed how, by solving analytically equation (\ref{eom33}) we can describe an unpinning wave that travels as a shock in the fluid. In doing so, however, we have had to make the assumption that $\gamma$, the fraction of unpinned vortices, is a constant, and neglect the steady state contribution to the vortex number density $\kappa n_\vv\approx 2 \Omega_\mathrm{n}$. In practice this allows us to only evolve the equations of motion as long as $\varpi\partial_\varpi\Omega_{\n}\ll 2\Omega$.

To go beyond this approximation we have to provide an evolution equation for the unpinned fraction $\gamma$. This is a complex problem, given that the micro-physics that governs vortex pinning acts on scales much smaller than the hydrodynamical scale. Attempts have been made in this direction in the context of superfluid turbulence, in which case an additional equation is included to model the evolution of the vortex length \citep{Mongiovi}.
Here, however, we will not derive a full mean-field description of vortex unpinning, but rather focus on how different prescriptions for short-timescale movement of vortices during an avalanche can affect astrophysical observables. Another approach to constructing global models was taken in \citep{fulg}, where the angular velocity lag between the pulsars's superfluid interior and a rigid crust is considered as fluctuating according to a state-dependent Poisson process. In this case local vortex motion and knock-on effects are not taken into account in the hydrodynamics.

In analogy with mean-field approaches to the study of sand-piles and of self organised critical systems more generally, we can assume that one has a time evolution equation for $\gamma$, with local terms which depend only on powers of $\gamma$ itself, that govern the long scale relaxation of the system close to equilibrium, due to random unpinning and repinning. 
The evolution equations for $\gamma$ thus take the form:
\be
\frac{\partial\gamma}{\partial t}=\sum_n \alpha_n \gamma^n+\xi f(\gamma, \partial_r\gamma)\label{alpha1}\;,
\ee
where $\alpha_n$ and $\xi$ are coefficients, and $f(\gamma, \partial_r\gamma)$ is a function of $\gamma$ and its spatial derivatives that models transport of vortices. In standard SOC models $f$ would simply be a diffusion term of the form $f=\nabla^2 \gamma$. In our case the Magnus force sets a preferred direction for vortex motion, so we will consider different forms of advection terms rather than diffusion.
The terms $\sum_n \alpha_n \gamma^n$ model the competition between unpinning and repinning on the microscopic level, and set the steady state equilibrium of the system. The addition of noise to (\ref{alpha1}) can then lead to departures from equilibrium and avalanches. This is, however, a complex problem, well beyond the scope of the current analysis. Here we intend to take a first step towards the analysis of the propagation of avalanches, therefore we will neglect the terms $\sum_n \alpha_n \gamma^n$, that set the background equilibrium on timescales longer that those of an avalanche and simply analyse how a large perturbation propagates by considering an evolution equation of the form:
\be
\frac{\partial\gamma}{\partial t}=\xi f(\gamma, \partial_r\gamma)\label{alpha2}.
\ee

\subsection{Vortex advection}

Let us consider different setups to describe the evolution of the unpinned fraction $\gamma$, neglecting the entrainment for simplicity. Given the phenomenological nature of this investigation we consider three setups.



\begin{figure}
\includegraphics [width = \linewidth]{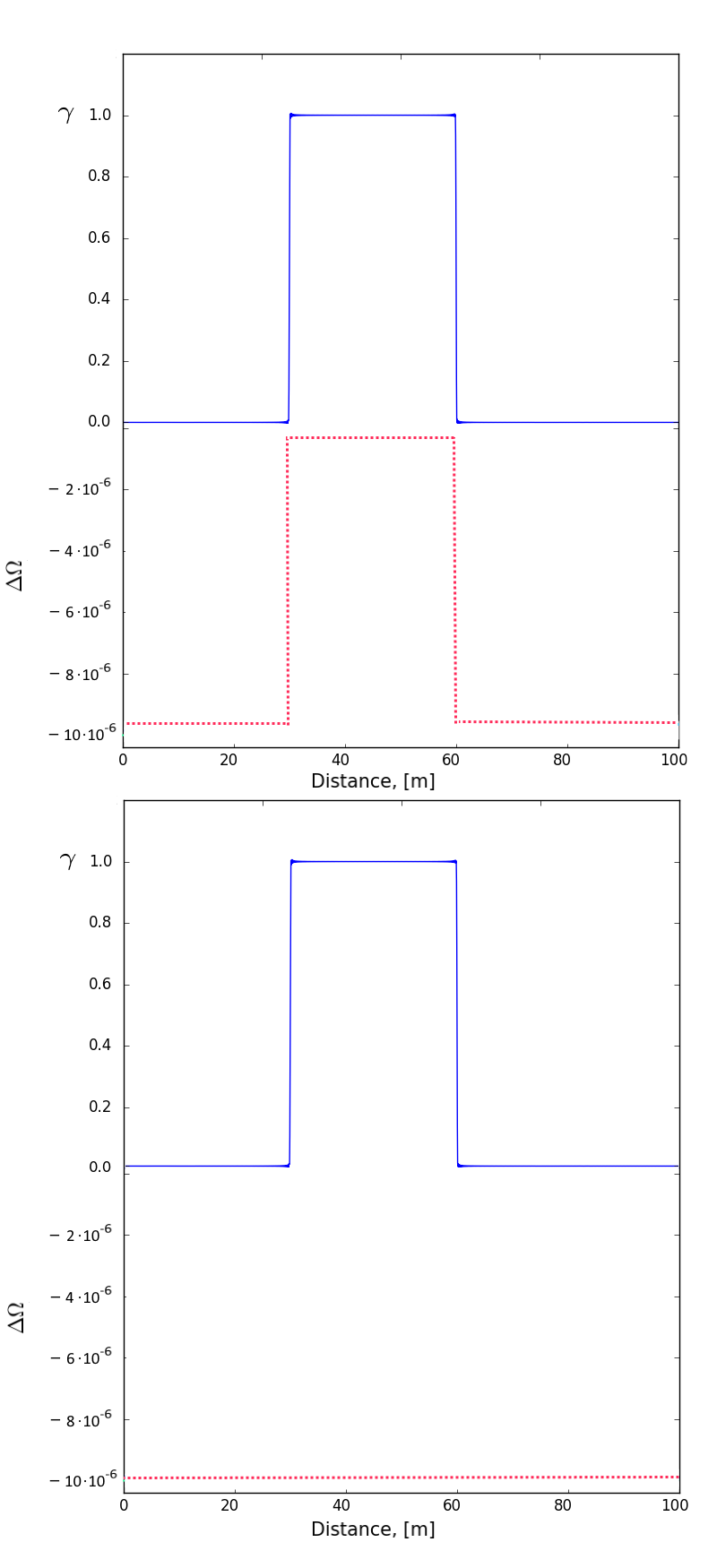}
\caption{
	Examples of initial conditions for step profiles both in $\Delta\Omega$ and $\gamma$ (top), as used in case III simulations,  and those with a flat profile in $\Delta\Omega$ and step in $\gamma$, used in case I and II to initiate a glitch (bottom). 
The lag $\Delta\Omega$ and the radial coordinate $\varpi$ are measured in $rad/s$ and in meters respectively.	
}
\label {fig:init1}
\end{figure}

In the first setup we consider a non-linear problem with non-constant coefficient in the advection term so that we allow vortices to advect with the velocity that is equal to the velocity of the shock wave. This is done to synchronise the repining process and angular momentum exchange, and account for vortex unpinning in the front. We explicitly take into account the steady state contribution $\kappa n_\vv\approx 2 \Omega_\mathrm{n}$ in equation (\ref{eqkappa}), but assume that all vortices are unpinned in the front ($\gamma=1$) in order to decouple the advection term in the lag from the equations of motion for $\gamma$. Our first setup (case I) thus takes the following form for $\Delta\Omega$:
\be
\frac{\partial\Delta \Omega }{\partial t} 
= 
\frac{\mathcal{B}}{{x}_{\p}}{\varpi}_{s} \Delta \Omega\frac{\partial\Delta \Omega}{\partial \varpi }-2\Omega \Delta \Omega\frac{\gamma \mathcal{B} }{{x}_{\p}} \label{eom3}\;,
\ee
and the equation for $\gamma$ is:
\be
\frac{\partial \gamma }{\partial t} = \frac{\mathcal{B}}{2{x}_{\p}}{\varpi}_{s} \Delta \Omega\frac{\partial\gamma}{\partial \varpi } \label{eom4}.
\ee
Equations (\ref {eom3}) and (\ref {eom4}) form a system to solve for two independent variables, $\gamma$ and $\Delta \Omega$. An extension to this setup has been made by reducing $\mathcal{B}$ by a factor 10 in the first term on the right hand side of equation (\ref{eom3}), thus obtaining  
\be
\frac{\partial\Delta \Omega }{\partial t} = \frac{\mathcal{B}}{10 {x}_{\p}}{\varpi}_{s} \Delta \Omega\frac{\partial\Delta \Omega}{\partial \varpi }-2\Omega \Delta \Omega\frac{\gamma \mathcal{B} }{{x}_{\p}} \label{eom3a}\;,
\ee
which corresponds to slowing down the velocity of free vortices and longer term evolution of the lag, while leaving unaltered the short term dynamics leading to the glitch rise. This allows us to mimic the physical situation in which a fast rise is followed by an unpinning wave in which not all vortices are unpinned in the front, but still retain the numerical advantage of decoupling the advection terms in (\ref{eom3a}) and (\ref{eom4}). Essentially the velocity of the unpinning wave is reduced while that of the initial exponential rise is not.


In a second setup (case II) we allow advection of $\gamma$ with a constant velocity equal to the initial velocity of the shock wave in $\Delta\Omega$. In this case the equation describing the evolution of $\gamma$ and $\Delta\Omega$ are:
\be
\frac{\partial\Delta \Omega }{\partial t} = \frac{ \gamma \mathcal{B} }{{x}_{\p}}{\varpi}_{s} \Delta \Omega\frac{\partial\Delta \Omega}{\partial \varpi }-2\Omega \Delta \Omega\frac{\gamma \mathcal{B} }{{x}_{\p}} \label{E1}\;,
\ee
\be
\frac{\partial \gamma }{\partial t} = \frac{\mathcal{B}}{2{x}_{\p}}{\varpi}_{s}{\Omega}_{init}\frac{\partial\gamma}{\partial \varpi } \label{E2}\;,
\ee
where ${\Omega}_{init}$ is the (negative) initial value of the lag $\Delta \Omega$. 
This case is interesting because here $\gamma$ advects with a constant velocity that does not depend on spatial changes in lag. Decoupling these processes means that the exchange of angular momentum between normal and superfluid component does not affect the propagation of free vortices. Both in case I and II we provide as initial conditions a pulse in $\gamma$ and a flat profile in $\Delta\Omega$, as shown in fig.\ref {fig:init1}. As we shall see the linear terms in the equations of motion for $\Delta\Omega$, in the presence of an increase in $\gamma$ lead to an exponential rise and rapidly lead to a step in $\Delta\Omega$.

The third setup (case III) is closely related to case II with constant advection, but now the parameter $\gamma$ is explicitly included in the equation for the lag $\Delta\Omega$, which is thus coupled to the evolution of $\gamma$. In order to make the problem numerically tractable the linear term is also excluded and we have an equation in the same form as (\ref {24}). The initial conditions differ from the previous ones, as the absence of a linear term mean that we cannot trigger a glitch simply with an increase in $\gamma$. Rather, we provide an initial step profile for $\gamma$ and $\Delta \Omega$, meaning that the lag is already formed and we force vortices to move as a result of a previous, unspecified, unpinning event. Examples of these initial conditions are also shown in fig.\ref {fig:init1}. The equations to solve are:

\be
\frac{\partial\Delta \Omega }{\partial t} = \frac{\mathcal{B}}{{x}_{\p}}{\varpi}_{s} \gamma \Delta \Omega\frac{\partial\Delta \Omega}{\partial \varpi }\;,
\label{lol1}
\ee
\be
\frac{\partial \gamma }{\partial t} = \frac{\mathcal{B}}{2{x}_{\p}}{\varpi}_{s} {\Omega}_{init}\frac{\partial\gamma}{\partial \varpi } \;.
\label{lol2}
\ee

The three setups are summarised in table (\ref{setups}).
\begin{table}
\begin{tabular}{l}
\hline
case I:\\
\\
$\frac{\partial\Delta \Omega }{\partial t} = \frac{\mathcal{B}}{{x}_{\p}}{\varpi}_{s} \Delta \Omega\frac{\partial\Delta \Omega}{\partial \varpi }-2\Omega \Delta \Omega\frac{\gamma \mathcal{B} }{{x}_{\p}}$\\
$\frac{\partial \gamma }{\partial t} = \frac{\mathcal{B}}{2{x}_{\p}}{\varpi}_{s} \Delta \Omega\frac{\partial\gamma}{\partial \varpi }$\\
\hline
case II:\\
\\
$\frac{\partial\Delta \Omega }{\partial t} = \frac{\gamma \mathcal{B}}{{x}_{\p}}{\varpi}_{s} \Delta \Omega\frac{\partial\Delta \Omega}{\partial \varpi }-2\Omega \Delta \Omega\frac{\gamma \mathcal{B} }{{x}_{\p}}$\\
$\frac{\partial  \gamma }{\partial t} = \frac{\mathcal{B}}{2{x}_{\p}}{\varpi}_{s}{\Omega}_{init}\frac{\partial\gamma}{\partial \varpi }$\\
\hline
case III:\\
\\
$\frac{\partial\Delta \Omega }{\partial t} = \frac{\mathcal{B}}{{x}_{\p}}{\varpi}_{s}\gamma \Delta \Omega\frac{\partial\Delta \Omega}{\partial \varpi }$\\
$\frac{\partial \gamma }{\partial t} = \frac{\mathcal{B}}{2{x}_{\p}}{\varpi}_{s} {\Omega}_{init}\frac{\partial\gamma}{\partial \varpi }$\\
\hline
\end{tabular}
\caption{Summary of the three different prescriptions that are used to couple vortex motion (evolution of the unpinned vortex fraction $\gamma$) and angular momentum exchange (evolution of the lag $\Delta\Omega$).}\label{setups}
\end{table}

These pairs of equations for the evolution of the lag and the fraction of unpinned vortices for the considered setups have been solved using the Dedalus spectral code \citep{dedalus} as well as with a 2-order Godunov finite difference code. Tests of the numerical solution's accuracy have also been implemented for both codes by means of direct measurements of the velocity of a shock wave in test problems.

We assume a constant mutual friction parameter $\mathcal{B}$ and assume that the distance travelled by vortices is small compared to the radius $\varpi_s$, which we take constant. In a realistic case mutual friction will depend on density and composition, thus on $\varpi_s$, and will be due to different processes in the crust and core, thus depending strongly on the location of our simulation box in the star.

\begin{figure}[h]
\includegraphics [width = \linewidth]{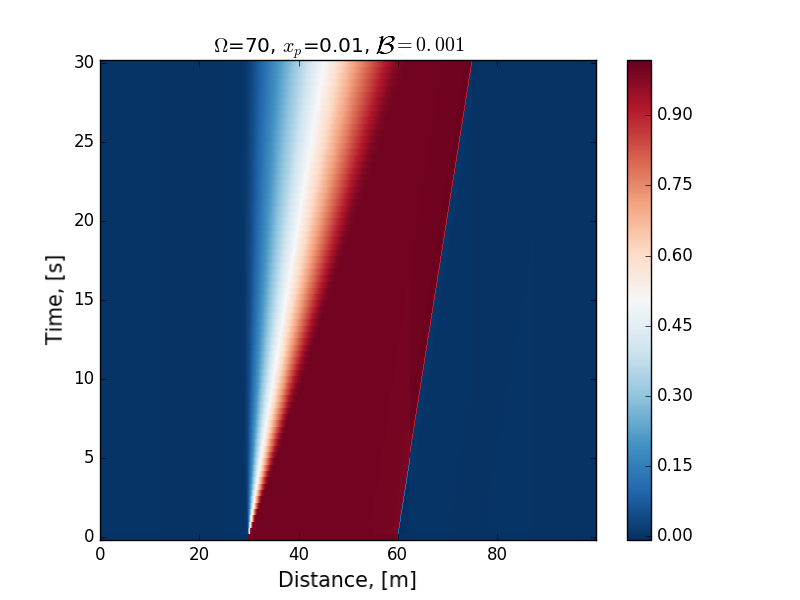}
\caption {Evolution of $\gamma$ for case I with step initial condition, as described in the text and seen in figure \ref{fig:init1}. The fraction of unpinned vortices decreases, thus approximating a repining process.}
\label {fig:f2}
\end{figure}

\begin{figure}[ht]
\includegraphics [width = \linewidth]{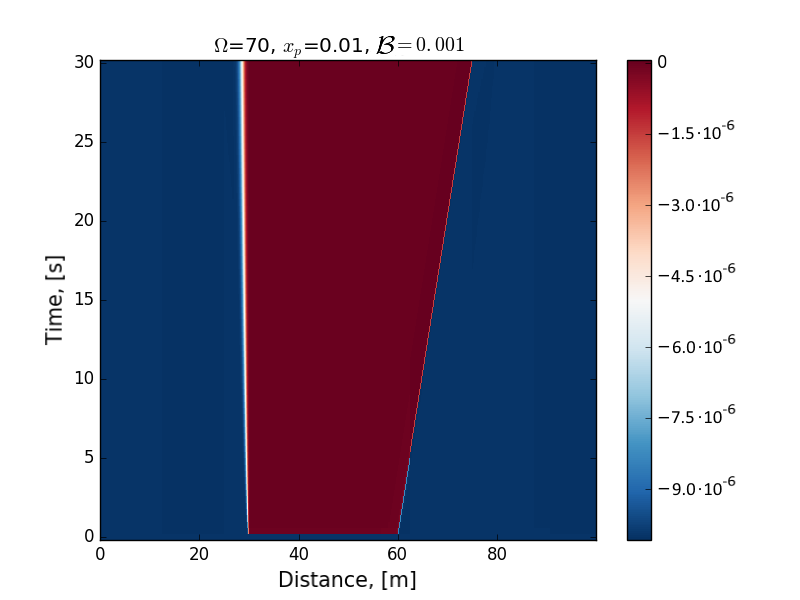}
\caption {Evolution of $\Delta \Omega$ for case I with step initial condition, as described in the text and seen in figure \ref{fig:init1}.}
\label {fig:f1}
\end{figure}

A typical evolution of $\Delta \Omega$ is shown in fig.\ref {fig:f1} for the synchronized velocity case (case I), while the evolution of $\gamma$ is shown in fig.\ref {fig:f2}. The pattern of the process is as follows: an initially flat profile creates a difference in angular velocities, due to the exchange of angular momentum between a normal and a superfluid component, and creates large gradients in $\Delta\Omega$. The decrease in lag then spreads out. The typical time for the rise in frequency is less than a second in our setup, but strongly depends on the poorly known mutual friction parameter $\mathcal{B}$. 

The evolution of the fraction of unpinned vortices is characterised by a decrease with time, which mimics vortex repining. Note that, in order to solve the equations numerically, artificial dissipation has been introduced.

Let us now turn our attention to the (internal) torque acting on the protons, i.e. on the `normal' component of the star that is coupled to the magnetic field and thus to the observable electromagnetic emission.
Locally the proton angular velocity evolves as:
\[
{\Omega }_{\p} = {\Omega }_{\n} + \Delta \Omega \, .
\]
In order to consider the motion of a rigid crust we will average over the interval by integrating between [${\varpi}_{1}$ , ${\varpi}_{2}$] that in our case represent the boundaries of our computational domain, i.e. minimum and maximum radial distance in a star at which the evolution of a system is simulated. Therefore, we define 
\[
\langle \Delta \Omega \rangle 
\, = \, 
\frac{1}{{\varpi}_{2}-{\varpi}_{1}}\int_{{\varpi}_{1}}^{{\varpi}_{2}} 
\Delta \Omega(\varpi, t) \, d\varpi \, .
\]
Note that in the above equation the lag has been averaged by using a uniform measure over the computational domain: in this initial calculation we neglect the effect induced by the non-uniform distribution of the moment of inertia inside the star and a more realistic calculation should retain this effect \citep{AP17}. 
The evolution of the charged component during a glitch triggered at $t=0$ is given by
\[
\Omega_p(t)  \, = \,\Omega_p(0) + x_n \, 
 \langle \Delta \Omega (t)\rangle 
- x_n  
\langle \Delta \Omega (0)\rangle  \, ,
\]
where $x_n$ is the ratio between the partial moment of inertia of superfluid neutrons and the total one.
Since we are not considering a realistic stratification of the stellar structure, in the following we show the evolution of the averaged lag $\langle \Delta \Omega \rangle$ instead of the charged component $\Omega_p(t)$: according to the above equation, they are qualitatively similar and share the same timescales.



\begin{figure}[h]
\includegraphics [width = \linewidth]{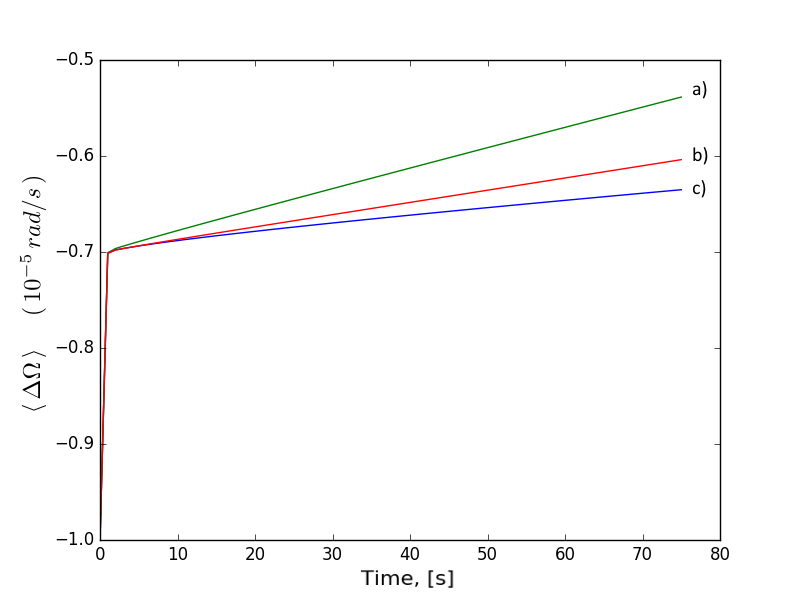}
\caption {
	Comparison of the evolution of the averaged lag $\langle \Delta\Omega \rangle$ over the computational box $[\varpi_1,\varpi_2]$ for the different setups. 
	a) Synchronized velocity setup (case I): [(\ref {eom3}), (\ref {eom4})] with an initially flat profile for $\Delta\Omega$ and step in $\gamma$; 
	b) Constant advection setup (case III)  [(\ref {lol1}), (\ref {lol2})] with initial step profiles in both $\Delta\Omega$  and $\gamma$; 
	c) Synchronized velocity (case I) with artificially decreased speed of free vortex propagation ($\mathcal{B}$ reduced as described in the text).}
\label {fig:f3}
\end{figure}

\begin{figure}[ht]
\includegraphics [width = \linewidth]{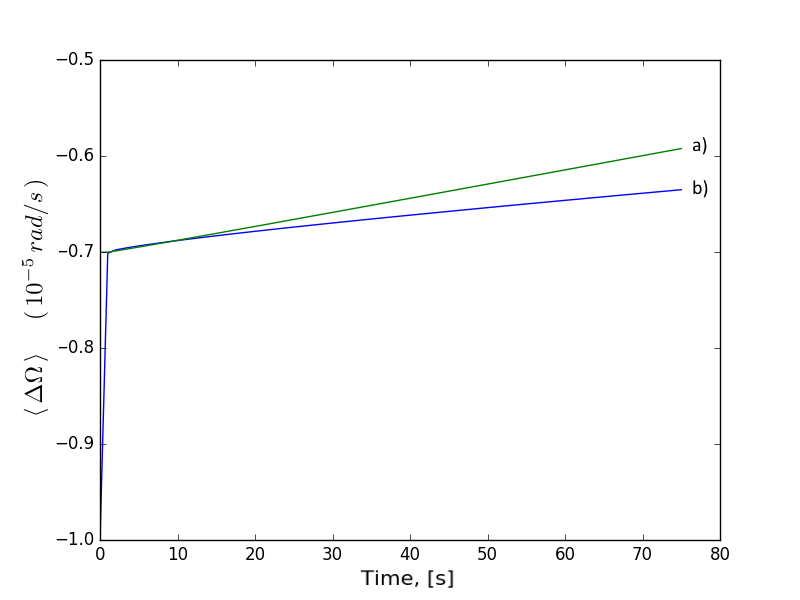}
\caption {
	Comparison of the evolution of the averaged lag $\langle \Delta\Omega \rangle$ for two setups with the constant advection term. 
	a) Constant advection (case II) [(\ref {E1}), (\ref {E2})]; 
	b) Constant advection of knocked on vortices (case III) [(\ref {lol1}), (\ref {lol2})]. We stress that in this plot, as well as in all the others involving $\langle \Delta \Omega \rangle $, the evolution of the charged component follows a similar curve with identical timescales.
}
\label {fig:f7}
\end{figure}

Using the previously obtained results for the lags $\Delta\Omega$ we average over the computational domain for each setup in order to obtain $\langle \Delta \Omega \rangle$; the results are shown in fig.\ref {fig:f3}. The comparison of the related scenarios for constant advection are shown in fig.\ref {fig:f7}. 

Generally, for the reference value $\mathcal{B}=10^{-3}$ as well as for the lower value $\mathcal{B}=10^{-4}$ the behaviour of the `normal' component is characterised by a rapid exponential rise, on timescales of seconds, followed by a slower, apparently linear, increase in frequency on timescales of a minute. This behaviour is, in fact, suggestive of what was observed for the 1989 glitch of the Crab pulsar, in which a fast and unresolved rise was followed by a slower component \citep{slowrise}.

As can be seen from  figures \ref{fig:f3} and \ref{fig:f7}, the rise time for the three major setups is almost the same, while there are differences in the linear responses. However the differences in frequency and frequency derivative are still small, and of order of $30\%$ for the frequency derivative, indicating that in the study of the short term rise and post-glitch behaviour the choice of setup does not strongly influence the conclusions.

\subsection{Glitch precursors}

We now discuss how differences in pinning strengths in the neutron star crust can influence the evolution of the frequency, and in particular whether unpinning in lower strength pinning regions can trigger unpinning and glitches in regions with stronger pinning and thus larger lags. In other words we are interested in examining whether smaller unpinning events, that may show up simply as changes in spindown rate, rather than steps in frequency, could be glitch precursors, as observed, for example, in the pulsar J0537-6910 \citep{Middle06}.

\begin{figure}[h]
\includegraphics [width = \linewidth]{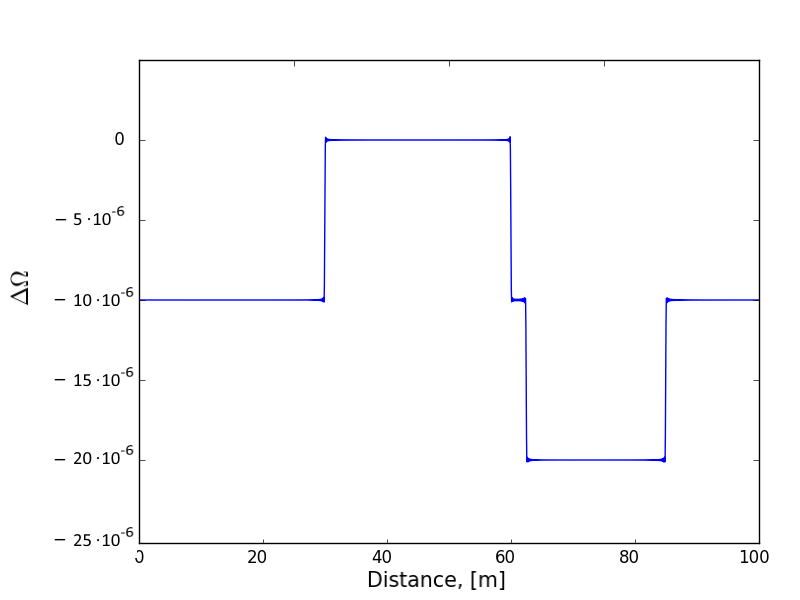}
\caption {
	Initial condition at $t=0$ for the lag $\Delta\Omega$ in units of $rad/s$ with a sequence of steps,  physically corresponding to different pinning strengths.
	The lag $\Delta\Omega$ and the radial coordinate $\varpi$ are measured in $rad/s$ and in meters respectively.}
\label {fig:init2}
\end{figure}

To do this let us consider the evolution of a system with the lag between the normal and the superfluid component that is not simply a step but a sequence of steps. Initial conditions for this case are shown on fig. \ref{fig:init2}. Different lags $\Delta\Omega$ correspond to a different pinning strength, as stronger pinning leads to a larger critical lag.

\begin{figure}[ht]
\includegraphics [width = \linewidth]{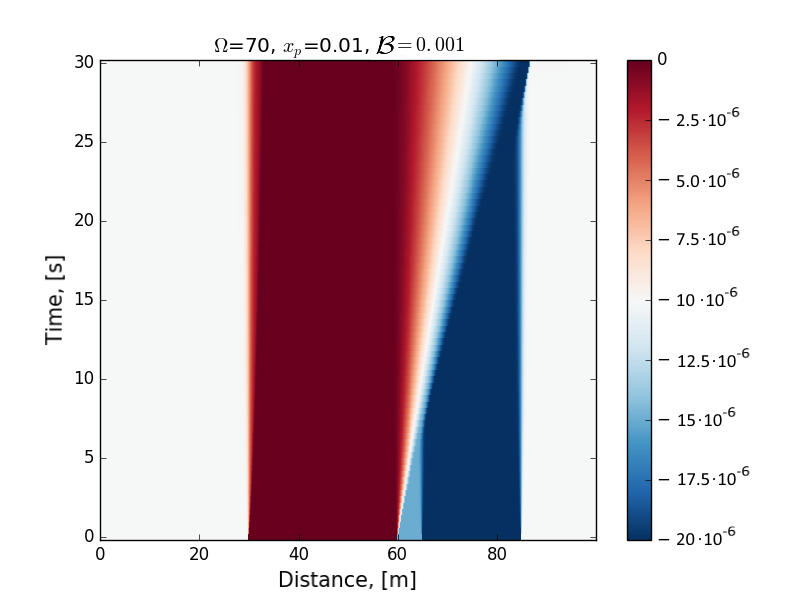}
\caption {Evolution of the lag $\Delta\Omega$ for the initial conditions in figure \ref{fig:init2}.} 
\label {fig:f666}
\end{figure}

\begin{figure}[ht]
\includegraphics [width = \linewidth]{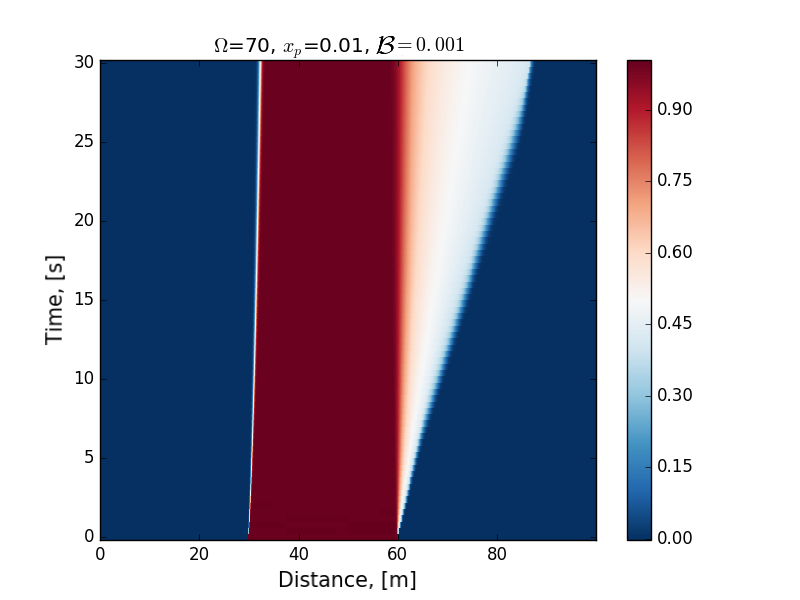}
\caption {Evolution of the parameter $\gamma$ for the initial conditions in figure \ref{fig:init2}.}
\label {fig:f667}
\end{figure}

To study the evolution we use the equations from case III, i.e. [(\ref {lol1}), (\ref {lol2})], with initial conditions for $\gamma$ being a step located at the same distance and with the same width as in fig.\ref {fig:init1}.  Note that the lag in the second region is twice that in the first (i.e. the pinning is twice as strong). Larger differences can be expected in neutron star crusts, but cannot be treated in our current numerical setup. The results of solving (\ref{lol1}) and (\ref{lol2}) are in fig.\ref{fig:f666} for the evolution of the lag $\Delta\Omega$, and in fig. \ref{fig:f667} for the evolution $\gamma$.

A propagating wave begins to travel due to the unpinned vortices. When reaching the region with a higher lag the transfer of angular momentum is much more effective and this results in increase of rotational rate as seen in the evolution of the charged component. Due to the presence of regions with a non uniform distribution of pinning forces, and thus of lag between the normal and the superfluid component, in real NSs a more complex pattern is likely to appear. The fraction of free vortices, in turn, goes through two transitions. The first transition occurs when free vortices reach the region with a higher lag, where the amount of free vortices that are able to continue moving further decreases. The second transition occurs when vortices pass this region. Their amount then decreases with a constant rate.

\begin{figure}[h]
\includegraphics [width = \linewidth]{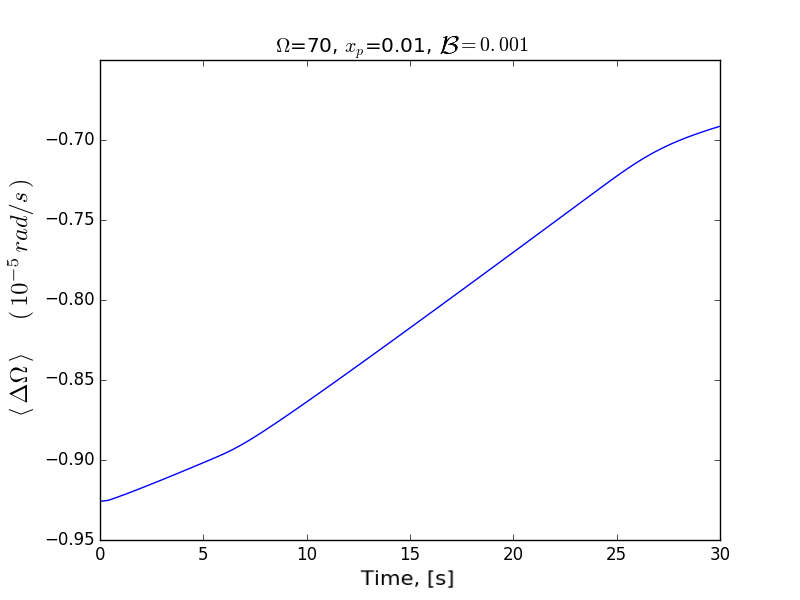}
\caption {Evolution of the averaged lag $\langle \Delta\Omega \rangle$ for initial conditions in figure \ref{fig:init2}. The difference in critical lags leads to an initial slower rise, followed by a faster increase in frequency when the unpinning front reaches the stronger pinning region. Even larger differences in critical lag (and equivalently pinning force) could lead to a faster and larger glitch after the initial precursor, but are numerically intractable in our setup.}
\label {fig:f777}
\end{figure}

The evolution of the averaged lag is shown in fig.\ref {fig:f777}. Unlike the other setups now the initial rise time is much longer. When free vortices reach the region with a higher lag the slope increases and decreases again after passing the region. This means that amplification of the initial rise is possible if in the outer region the pinning force is stronger.

As a result, the star's angular velocity may change not only abruptly, showing a glitch-like rise, but also more gradually, depending on local properties of the region, i.e. the distribution of areas with uneven pinning but also the local value of the mutual friction and the moment of inertia of the fluid. 
 In general our results indicate that an initial increase in the spin-down rate (decrease in absolute value) may be the precursor of a larger glitch, although the differences in pinning required for this scenario are larger than those that our numerical setup allows. We are thus unable to simulate physically realistic sizes and timescales.

This behaviour, however, is similar to what has been observed in pulsar J0537-6910 \citep{Middle06, Ferdman}, where the preglitch behaviour exhibits brief `upticks' and `downticks' in $\dot{\nu }$ of varying amplitudes and durations. The timescales are different from those that we simulate, due to our numerical limitations. However, our results, although they depend on poorly known physical quantities in the crust of the star, indicate that it is possible that the same process in a NS may lead to different phenomena. In the case presented here the interaction between the superfluid and the normal component gives a rise to a glitch precursor, while the same process in an isolated strong pinning region leads to a standard glitch. 

\subsection{Frequency decrease and anti-glitches}

\begin{figure}
\includegraphics [width = \linewidth]{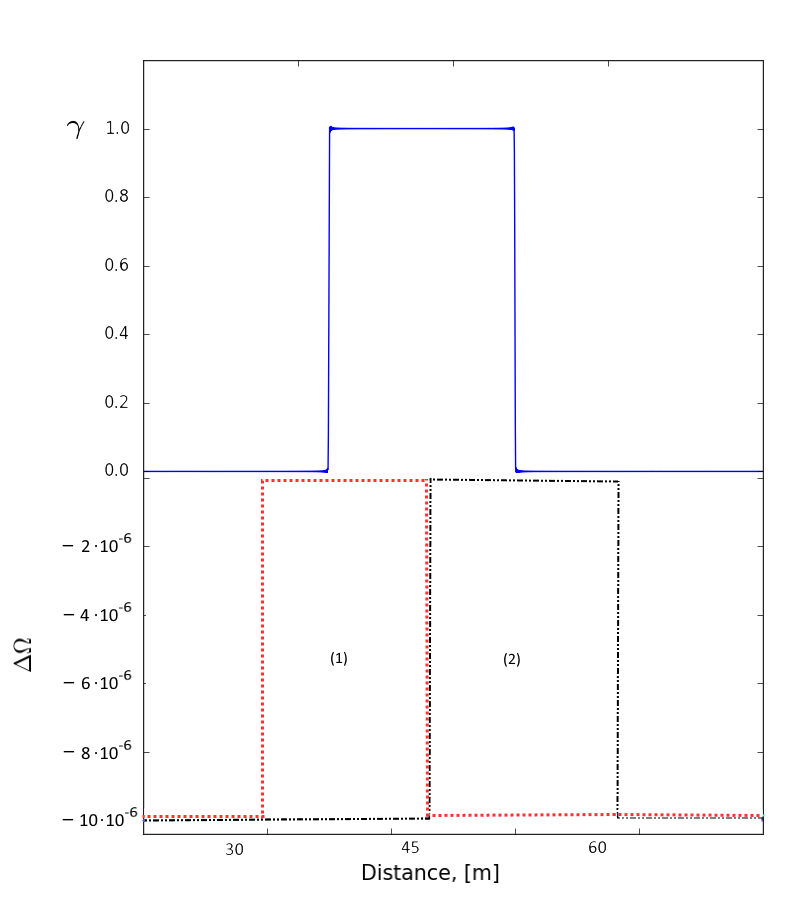}
\caption{
	Initial conditions for $\gamma$ and $\Delta\Omega$ for the anti-glitch test cases described in the text. Two cases are shown: (1) conditions leading to and increase of the angular velocity; (2) conditions leading to a decrease of the angular velocity, or anti-glitch.
    The lag $\Delta\Omega$ is measured in $rad/s$ and the radial coordinate $\varpi$ in meters.
	} 
\label {fig:init3}
\end{figure}

The non-linear evolution we consider can, however, lead to other surprising results in the framework of the standard glitch model.  In particular we find setups in which not only an increase, but also a decrease in a star's angular velocity can be obtained. To do this we study the evolution of a system with step initial conditions using the setup in case III, corresponding to equations (\ref{lol1}) and (\ref{lol2}). Initial conditions for two cases are shown in fig.\ref {fig:init3}.

\begin{figure}
\includegraphics [width = \linewidth]{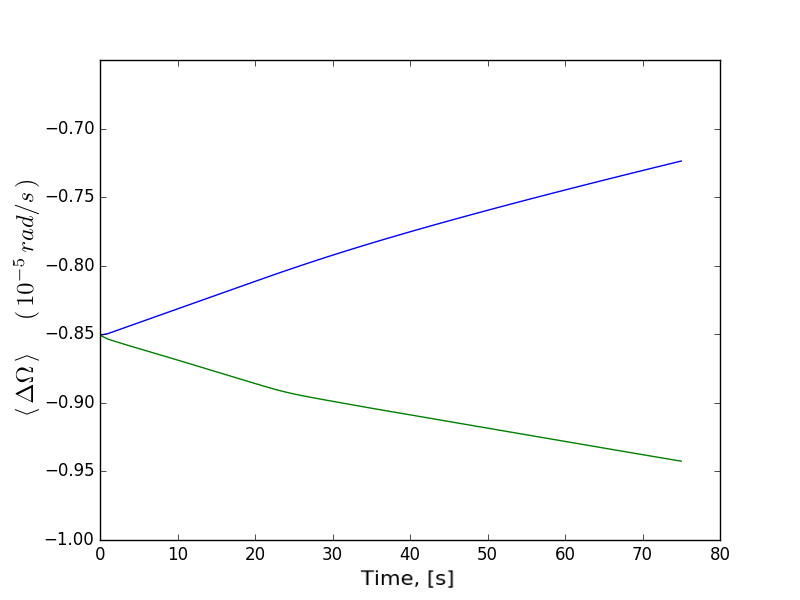}
\caption {
	Evolution in time of the averaged lag $\langle \Delta\Omega \rangle$  for a glitch and anti-glitch tests. The blue rising curve corresponds to the glitch-like rise, the green decreasing curve to the anti-glitch behaviour. Note that, in both cases, the evolution of the charged component follows similar curves with identical timescales.
	} 
\label {fig:plus}
\end{figure}

The difference between the two setups is that in the first case the region with null lag is located `behind' the region with free vortices and these regions partially coincide while in the second case the null-lag region is located further out than the front. Results for this case are shown in fig.\ref {fig:plus}. 

As can be seen from the figure, unpinned vortices in the area behind the zero lag region start to exchange angular momentum, which tends to increase the angular velocity. However propagation tends to decrease the extent of the coupled region at a faster rate, resulting in an `anti-glitch', i.e. a local decrease of angular velocity. This behaviour is intriguing, given observations of such anti-glitches in magnetars \citep{antiglitch}. However, since the decrease in frequency is the result of a competition of process, it is possible that the overall anti-glitch behaviour is the consequence of our particular setup. As for the influence of the numerical dissipation, several test have been made in order to study the changes of the anti-glitch behaviour appearance as well as the time of rise. It was found that it has minimal impact on the results, and the feature is robust for our setup. However further investigation in a more realistic scenario will be required to determine whether this evolution is physically significant.

Other initial conditions, in fact, result in much more predictable evolutions and show an increase of angular velocity, as shown in fig.\ref {fig:plus} as a blue curve, while the anti-glitch is shown as green curve.

\begin{figure}
\includegraphics [width = \linewidth]{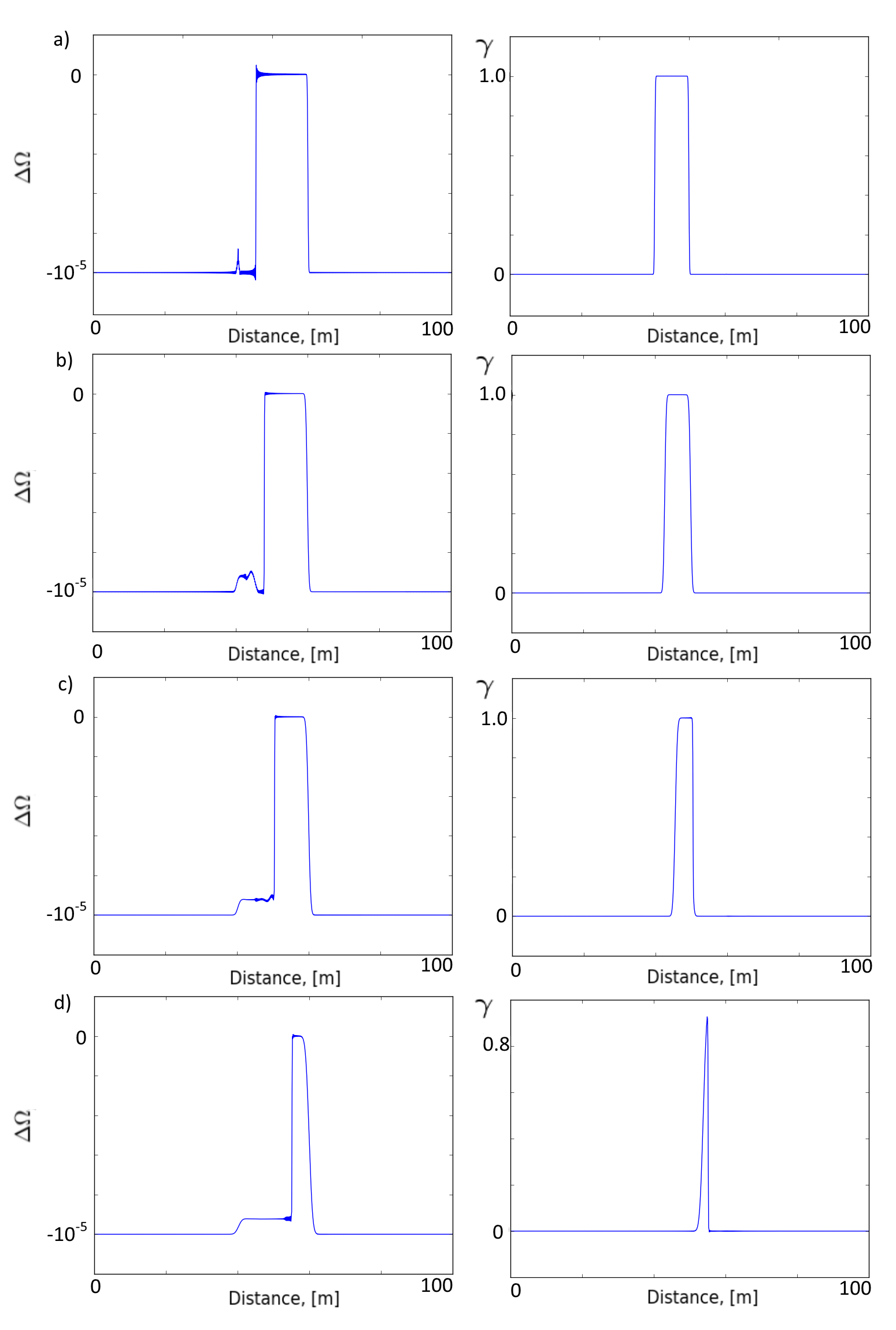}
\caption {Snapshots of evolution in time for $\Delta\Omega$ (left column) and $\gamma$ (right column) a) After 1 sec; b) after 6 sec; c) after 12 seconds; d) after 30 seconds. Here the lag $\Delta \Omega$ is measured in $rad/s$ and the radial coordinate $\varpi$ in meters.} 
\label {fig:snapshots}
\end{figure}

Let us study in detail the evolution of the solution where the anti-glitch appears. For this we show four snapshots for $\gamma$ and $\Delta\Omega$ on fig.\ref {fig:snapshots}. 

Initially unpinned vortices form a small peak in $\Delta\Omega$ behind the main region of the lag. This peak grows with time until it reaches the main region with zero lag. Next these two regions start to interact, forcing the initial region with vortices to decrease in size and move. Further interaction makes both regions move, $\gamma$ decrease and an unpinning wave propagate. The overall outcome, whether an increase or decrease in frequency, is thus sensitive to the timescale on which these processes occur and the speed of propagation of $\gamma$.

\section {Parameter study}
To study the influence of the different parameters on the evolution of the solution we first change the mutual friction, i.e. the $\mathcal{B}$ parameter. In a realistic star this parameter depends on density, and will thus depend on the location of the computational box in the NS.  Since we take the mutual friction to be constant it represents the averaged value over the computational domain, i.e. over the path of vortex movement. Decreasing the mutual friction will generally increase the timescale for the rise, while higher mutual friction leads to a faster glitch, for a fixed initial setup.

\begin{figure}
\includegraphics [width = \linewidth]{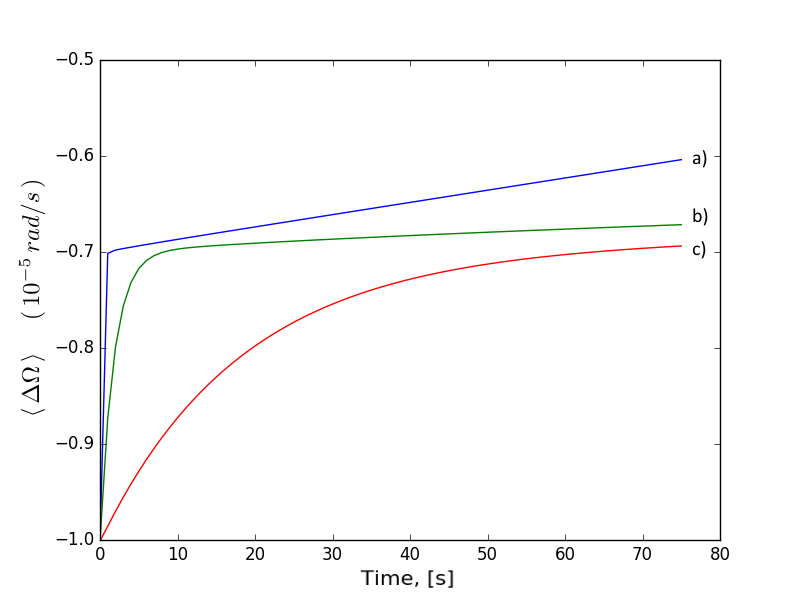}
\caption {
	Evolution in time of the averaged lag $\langle \Delta\Omega \rangle$  - Influence of the mutual friction parameter $\mathcal{B}$ on the speed of rise for case II. a) $ \mathcal{B} = {10}^{-3}$; b) $ \mathcal{B} = {10}^{-4}$ ; c) $ \mathcal{B} = {10}^{-5} \,$.} 
\label {fig:f4}
\end{figure}

\begin{figure}
\includegraphics [width = \linewidth]{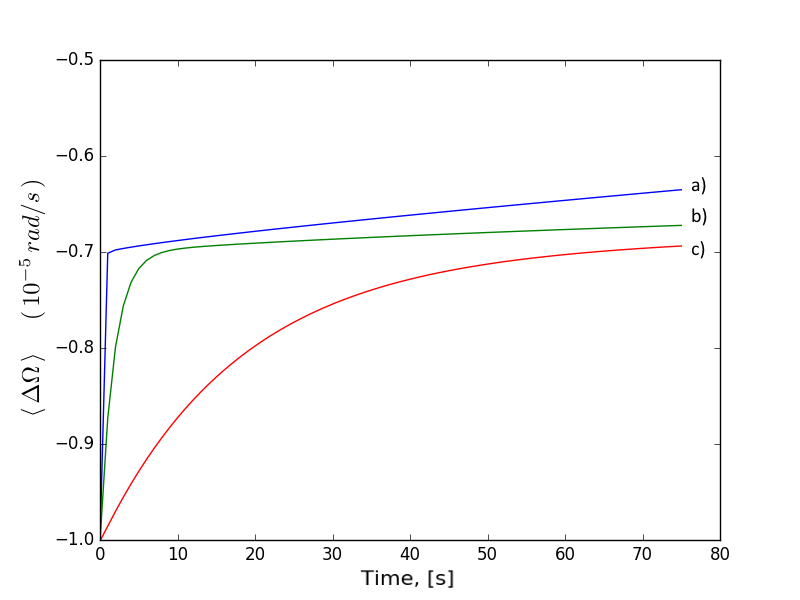}
\caption {
	Evolution in time of the averaged lag $\langle \Delta\Omega \rangle$  - Influence of the mutual friction parameter $\mathcal{B}$ on the speed of rise for case I. a) $  \mathcal{B} = {10}^{-3}$; b) $  \mathcal{B} = {10}^{-4}$; c) $  \mathcal{B} = {10}^{-5} \,$.}
\label {fig:f5}
\end{figure}

The dependance is intuitively correct since the mutual friction is the mechanism that is responsible for an angular's momentum transfer strength, it's a `bridge' between the normal and the superfluid component, and the behaviour can easily be understood from equation (\ref{tscale}) for the coupling timescale between components, if we neglect non-linear terms. The consequences of the mutual friction variations are shown in fig.\ref {fig:f4} for case II as a representative of a non-constant advection family and in fig.\ref {fig:f5} for case I with constant advection. 

Note that $\gamma\mathcal{B}$ is the parameter that affects the character of a glitch. Its evolution leads, for example to different kinds of  relaxation even in a single pulsar, given that in a realistic system it is not constant in time (due to vortex pinning and unpinning) or constant along the path of vortex movement \citep{HasAnt}.

\begin{figure}
\includegraphics [width = \linewidth]{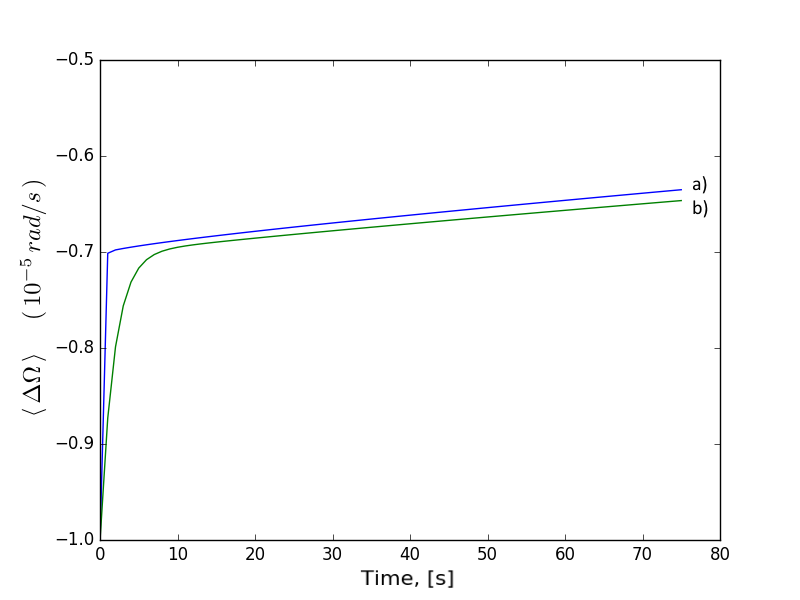}
\caption {
	Evolution in time of the averaged lag $\langle \Delta\Omega \rangle$  - Influence of the angular frequency $\Omega$ of a star on the evolution for constant advection, i.e. case I: a) $\Omega=70  \, rad/s$, b) $\Omega=7 \, rad/s$. }
\label {fig:f6}
\end{figure}

Next let us study the influence of the angular frequency of a star. In all of the simulations the angular frequency is initially equal to 70 $rad/s$ which is approximately equal to the Vela pulsars's angular velocity. In order to see how the unpinning wave propagation reacts we experiment with changing it to 7 $rad/s$. Results are shown on fig.\ref {fig:f6}. As expected from the linear analysis in (\ref{tscale}) decreasing the angular velocity of a star increases the rise time but does not strongly affect the results in the non-linear regime.

Changing the proton fraction  ${x}_{\p}$  acts as a simple rescaling of the mutual friction parameter $\mathcal{B}$ in our simple setup, and thus does not significantly affect the results for reasonable values of the parameter in the crust.

\section{Conclusions}

In this paper we have outlined a formalism for simulating the motion of superfluid vortex over-densities and fronts in hydrodynamical two-fluid simulations of pulsar glitches. We have shown that accounting explicitly for the differential rotation that is built up due to vortex accumulation introduces additional non-linear terms in the evolution equations for the lag $\Delta\Omega$, which allow for travelling waves (`unpinning' waves) as solutions.
The observational consequence of this setup is that coupling between the normal and superfluid components of the neutron star, mediated by vortex motion, can lead not only to exponential terms in the evolution of a pulsar's frequency, but also linear terms, and generally slower, longer term variations in the spin frequency of a neutron star.

We take our model one step further and introduce an additional parameter to the evolution, namely the fraction of free vortices in the system, $\gamma$. This parameter encodes the sub-grid physics of vortex motion that is not resolved on a hydrodynamical level, and depends on the complex quantum mechanical statistical processes that govern vortex interactions at a microscopic level. While micro-physical simulations have been relatively successful in investigating the relation between pulsar glitches and self organised criticality \citep{Lila11, HasRev} in small systems, or pinning of a vortex to a single defect \citep{Wlaz16}, they are still not at the level where contact can be made with large scale descriptions. We thus propose three phenomenological models for vortex motion, all of which mimic advection of free vortices together with propagating fronts in the lag between the normal component and the superfluid neutrons.

We study the evolution of the lag $\Delta\Omega$ and the free vortex fraction $\gamma$ in several setups for all three our prescriptions for varying mutual friction parameters $\mathcal{B}$ and rotation rates $\Omega$.
The main conclusion is that localised unpinning leads to an initial rapid rise, on the timescale of seconds or less for mutual friction parameters $\mathcal{B}>10^{-4}$, as one may expect due to electron scattering of magnetised vortex cores in the presence of superconducting protons \citep{AlparMF}, or due to Kelvin waves as the vortices move past nuclear clusters in the crust \citep{JonesKelvon, EB92}. This phase is, however, generally followed by a slower, quasi-linear rise on timescales of a minute, which is similar to what was observed in the 1989 glitch of the Crab pulsar \citep{slowrise}.
Overall the prescription we use for motion of the vortex fraction has little influence on the exponential rise, which is mainly due to the linear terms, but impacts on the slower long term evolution. Nevertheless the evolution of the normal component frequency is qualitatively similar, with only modest differences in rotational rates and frequency derivatives, between the three cases, which gives us confidence that our conclusions are robust and do not depend strongly on how we approximate the sub-grid physics of vortex motion.

We have also investigated how changes in pinning strength, approximated by different initial conditions for the lag, can impact the evolution of the frequency and the glitch.  We find that if there are regions in which pinning decreases with density, as one expects in the deep crust \citep{Seveso16}, then an initial unpinning event may lead to a slow change in frequency as a precursor of a larger glitch, triggered when the unpinning front reaches the stronger pinning region. Such precursor events may, in fact, have been observed before a number of glitches in pulsar J0537-6910 \citep{Middle06, Ferdman}, where `upticks' and `downticks' in $\dot{\nu }$ of varying amplitudes and durations were observed prior to several glitches.

We also find specific setups in which vortex motion can lead to a decrease in frequency, or an anti-glitch, such as that observed in the magnetar 1E 2259+586 \citep{antiglitch}. This behaviour is intriguing, as it would provide an explanation for this phenomenon in the standard glitch model (see also \citet{GK14} for an alternative approach). In our setup the feature is robust to changes in numerical dissipation, and does not appear to be a numerical artefact. Nevertheless a more detailed study in a more realistic setup is necessary to understand whether such an evolution is physically significant and would occur in a neutron star.

Despite the uncertainties, both due to the implementation of vortex motion, and poorly constrained physical parameters in the interior of the neutron stars, our simple models highlight the importance of allowing for vortex motion and accumulation in hydrodynamical simulations, as this allows for new and qualitatively different behaviour before, during and after a glitch.
On the other hand, our models are also further confirmation that the large scale response of the star strongly impacts on conclusions drawn from small scale vortex dynamics alone, as was already shown to be the case for size and waiting time distributions \citep{Haskell16}.
Future work should thus focus on further bridging the gap in scales between microscopic quantum mechanical simulations of vortex motion and large scale hydrodynamical models of superfluid neutron stars.

\section{Acknowledgements}

We acknowledge support from the Polish National Science Centre (SONATA BIS 2015/18/E/ST9/00577, P.I.: B.Haskell) and from the European Union's Horizon 2020 research and innovation programme under grant agreement No. 702713. Partial support comes from NewCompStar, COST Action MP1304.

We thank Varadarajan Parthasarathy, Morgane Fortin and Marco Antonelli for constructive discussions.

\bibliographystyle{mnras}

\bibliography{glitches}


\label{lastpage}

\end{document}